\documentclass[graybox]{svmult}

\usepackage{helvet}         
\usepackage{courier}        
\usepackage{type1cm}        
\usepackage{makeidx}         
\usepackage{graphicx}        
\usepackage{multicol}        
\usepackage[bottom]{footmisc}
\usepackage{amsmath,amsfonts}
\usepackage{amssymb}
\newcommand{\beqa}{\begin{eqnarray}}
\newcommand{\eeqa}{\end{eqnarray}}

\newcommand{\nn}{\nonumber}
\newcommand{\g}{\mathfrak{g}}

\begin{document}

\title*{Kac-Moody and Virasoro algebras on the two-sphere and the two-torus}
\author{Rutwig Campoamor-Stursberg and Michel Rausch de Traubenberg}
\institute{Rutwig Campoamor-Stursberg \at  Instituto de Matem\'atica Interdisciplinar and Dpto. Geometr\'\i a y Topolog\'\i a, UCM, E-28040 Madrid, Spain \email{rutwig@ucm.es}
\and Michel Rausch de Traubenberg \at  Universit\'e de Strasbourg, CNRS, IPHC UMR7178, F-67037 Strasbourg Cedex, France \email{Michel.Rausch@iphc.cnrs.fr}}
\maketitle

\abstract{ We particularise the construction of generalised Kac-Moody algebras associated to compact real
manifolds to the case of the two-torus $\mathbb T_2$ and the two-sphere ${\mathbb S}^2$. It is shown that these
algebras, as well as a Virasoro algebra associated to
the two-torus and the two-sphere, can also be derived considering the usual Kac-Moody and Virasoro algebras.}

\section{Introduction}
\label{sec:1}

Soon after their original formulation, Kac-Moody algebras \cite{Kac,Moo}, as well as their associated Virasoro algebras, were recognised to constitute a powerful tool in the context of symmetry unification and the study of locality properties in several physical problems \cite{Dol}. These constructions have motivated several generalisations, mostly based on analytical procedures, and complementary to the axiomatic approach \cite{Mdo,RS,RS2}, and where the prominent role played by the one-dimensional sphere $\mathbb S^1$ has been emphasised. This fact has suggested to look for generalisations of the procedure for higher dimensional either compact or general manifolds, for which additional criteria to ensure the convergence of integrals are given \cite{RMM}. This approach, besides its mathematical or geometrical interest, constitutes a possible path for extending the description of physical phenomena to theories formulated in higher dimensional space-times, where the Kaluza-Klein formalism or the supergravity theories may be the most prominent candidates for such generalisations.

\section{Kac-Moody algebras associated to compact manifolds}
\label{sec:2}

In \cite{RMM} and \cite{RMM2} it was shown that to any Lie algebra  $\mathfrak{g}$ (for simplicity, we assume that $\g$
is the compact real form of a complex simple Lie algebra) one can define a generalised Kac-Moody algebra associated to a compact real
manifold ${\cal M}$, denoted by $\widehat{\mathfrak{g}}({\cal M})$. In this short note ( for details on the algorithm leading to  $\widehat{\mathfrak{g}}({\cal M})$, the reader is referred to the previous references) we focus on the simplest  manifold, namely, when ${\cal M}$ is either the two-torus $\mathbb T_2$ or the two-sphere ${\mathbb S}^2$.  Kac-Moody algebras associated to these manifolds have been studied by several authors from different perspectives \cite{bars, MRT,HK-T, FRS,HH}.

\subsection{Kac-Moody algebras associated to $\mathbb T_2$ and ${\mathbb S}^2$}
\label{sec:KM}

Let $f^{ab}{}_c$ denote the structure constants of $\mathfrak{g}$, while $\delta^{ab}$ denotes the Killing form.

The Kac-Moody algebra associated to the two-torus, also called toroidal algebra in \cite{MRT}, is generated
by
\beqa
\widehat{\mathfrak{g}}(\mathbb T_2) = \Big\{T^a_{mn}, \,a=1,\cdots, \dim \mathfrak{g},\, m,n\in \mathbb Z, \,d_1, d_2, k_1, k_2\Big\} \nn
\eeqa
with non-vanishing Lie brackets:
\beqa\label{eq:torus}
\begin{split}
\big[T^a_{mn},T^b_{pq}\big]=& {\rm i} f^{ab}{}_c T^c_{m+p n+q} + (k_1 m + k_2 n)\delta^{ab} \delta_{m+p} \delta_{n+q} \ ,\\
\big[d_1,T^a_{mn}\big]= & m T^a_{mn},\qquad
\big[d_2,T^a_{mn}\big]= n T^a_{mn} \ ,
\end{split}
\eeqa
 where ${\rm i}=\sqrt{-1}$. It can be easily seen that $\widehat{\mathfrak{g}}(\mathbb T_2)$ possesses two Hermitian operators $d_1,d_2$ that correspond to the dimension of the two-torus. The generators $k_1,k_2$ are central extensions, in duality with $d_1,d_2$ ({ see \cite{RMM} for details).

 For the two-sphere, the associated Kac-Moody algebra is generated by
\beqa
\widehat{\mathfrak{g}}(\mathbb S^2) = \Big\{T^a_{\ell m}, \,a=1,\cdots, \dim \mathfrak{g},\, m \in \mathbb Z,
\,\ell \ge |m|, \,d, k\Big\} \nn
\eeqa
and the non-vanishing Lie brackets are given by
\beqa
\label{eq:sphere}
\Big[T^{a_1}_{\ell_1 m_1}, T^{a_2}_{\ell_2m_2} \Big]&=&{\rm i} f^{a_1 a_2}{}_{a_3} c_{\ell_1,m_1,\ell_2,m_2}{}^{\ell_3,m_3} T^{a_3}_
    {\ell_3 m_3}+
(-1)^{m_2} k m_1\delta^{ab} \delta_{\ell_1 \ell_2} \delta_{m_1+m_2}\nn\\
\Big[d,T^a_{\ell m}\Big] &=& m  T^a_{\ell m} \ ,
\eeqa
where
\beqa
c_{\ell_1,m_1,\ell_2,m_2}{}^{\ell_3,m_3} = \sum\limits_{\ell_3 = |\ell_1-\ell_2|}^{\ell_1+\ell_2} \sqrt{\frac  {(2\ell_1 +1)(2\ell_2 +1)} {2\ell +1}}  \Big({}^{\ell_1}_{0} \ \ {}^{\ell_2}_{0} \Big| {}^{\ell}_{0}  \Big)
\Big({}^{\ell_1}_{m_1} \ \ {}^{\ell_2}_{m_2} \Big| {}^{\hskip .45truecm \ell_3}_{m_1+m_2}  \Big)\ \nn
\eeqa
with
$
\Big({}^{\ell_1}_{m_1} \ \ {}^{\ell_2}_{m_2} \Big| {}^{\hskip .45truecm \ell_3}_{m_1+m_2}  \Big)\nn
$
denote the Clebsch-Gordan coefficients of $SO(3)$ \cite{RMM}.
The algebra $\widehat{\mathfrak{g}}(\mathbb S^2)$ only admits one Hermitian operator $d$,  whilst
$\mathbb S^2$ is two-dimensional,  implying that there is an important difference, when compared to  $\widehat{\mathfrak{g}}(\mathbb T_2)$
(see in particular Sec. \ref{sec:root}). The generator $k$ is a central extension, in duality with $d$.
Note that when the compact manifold ${\cal M}$  is either a compact Lie group $G_c$ (not to be confused with $G$, the Lie group associated to $\g$) or a coset space $G_c/H$ with $H$ a subgroup of $G_c$,  as a consequence of the Peter-Weyl theorem, the structure constants of $\widehat{\mathfrak{g}}({\cal M})$
always involve the Clebsch-Gordan coefficients
of the manifold $G_c$ \cite{RMM}.

The two algebras \eqref{eq:torus} and \eqref{eq:sphere} have different structural properties. Indeed, for the two-torus
the subalgebra
\beqa
\label{eq:aff}
\{T^a_{m,0}, a=1,\cdots,\dim \g, m\in \mathbb Z, d_1,k_1\}\subset \widehat{\g}(\mathbb T_2)
\eeqa
is isomorphic to the affine Lie algebra $\widehat{\g}$ associated to $\g$ (in our language, we could write  $\widehat{\g} =\widehat{\g}(\mathbb S^1)$, {\it i.e.},  $\widehat{\g}$ is the algebra associated
to the circle $\mathbb S^1$). In the case of the two-sphere,  it is not possible to identify a subalgebra
isomorphic to the affine Lie algebra.
\subsection{Root structure}
\label{sec:root}
The algebras introduced so far, {\it i.e.},
 Kac-Moody algebras associated to
 $\mathbb S^2$ and $\mathbb T_2$, can be seen as
a natural extension of affine Lie algebras, since affine Lie algebras
are related to the circle $\mathbb S^1$.
However, there exist other extensions of affine Lie algebras, called Kac-Moody
algebras, defined by means of  a system
of simple roots or a generalised Cartan matrix \cite{Kac, Moo, Mdo}. Thus, a comparison of Kac--Moody
algebras associated to  generalised Cartan matrices and the algebras introduced in Sec. \ref{sec:KM} is necessary.

Suppose that the Lie algebra $\mathfrak g$ has rank $r$. Introduce a system of simple roots $\alpha_{(1)},\cdots,\alpha_{(r)}$ and let $\Sigma$ denotes the root system of $\g$. Then,  we have $\mathfrak{g}=\big\{H^i, i=1,\cdots,r,
E_\alpha, \alpha \in \Sigma\}$ with $H^i,i=1,\cdots, r$ the generators of the Cartan subalgebra and $E_\alpha, \alpha\in\Sigma$ the generators associated to the roots $\alpha$.

With this decomposition of $\mathfrak g$, the Kac-Moody algebras associated to the two-torus
\eqref{eq:torus} takes the form \cite{RMM}:
\beqa
\label{eq:KM-S1xS1}
\big[H^i_{m_1,m_2}, H^j_{n_1,n_2}\big]&=&(k_1 m_1 +k_2 m_2) \delta^{ij} \delta_{m_1 + n_1} \delta_{m_2 + n_2} \ , \nn\\
\big[H^i_{m_1,m_2}, E_{\alpha,n_1,n_2}\big]&=& \alpha^i E_{\alpha,m_1+ n_1,m_2+n_2} \ , \\
\big[E_{\alpha,m_1,m_2}, E_{\beta,n_1,n_2}\big] &=& \left\{\begin{array}{ll}
\epsilon(\alpha,\beta) E_{\alpha+\beta, m_1+n_1,m_2+n_2}&\text{if} \ \ \alpha+ \beta\in \Sigma\\
\alpha\cdot H_{m_1+n_1,m_2+n_2} +\\ (m_1 k_1 +m_2 k_2) \delta_{m_1 + n_1} \delta_{m_2 + n_2}   &\text{if} \ \ \alpha + \beta=0\\
0&\text{otherwise}
\end{array}
\right.\nn\\
\big[d_i,H^j_{m_1,m_2}\big]&=& m_i H^j_{m_1,m_2} \ , \ \
\big[d_i,E_{\alpha,m_1,m_2}\big]= m_i E_{\alpha,m_1,m_2} \ , \ \  i=1,2\ . \nn
\eeqa
For the two-sphere \eqref{eq:sphere} the algebra reduces to  (see \cite{RMM}):
\beqa
\label{eq:KM-S2}
\big[H^i_{\ell_1,m_1}, H^j_{\ell_2,m_2}\big]&=& (-1)^{m_1} k m_1 \delta^{ij} \delta_{\ell_1,\ell_2} \delta_{m_1 + m_2}   \ , \nn\\
\big[H^i_{\ell_1,m_1}, E_{\alpha,\ell_2, m_2 }\big]&=& c_{\ell_1,m_1,\ell_2,m_2}{}^{\ell_3,m_3} \alpha^i E_{\alpha,\ell_3, m_3} \ , \\
\big[E_{\alpha,\ell_1,m_1}, E_{\beta,\ell_2,m_2}\big] &=& \left\{\begin{array}{ll}
\epsilon(\alpha,\beta) c_{\ell_1,m_1,\ell_2,m_2}{}^{\ell_3,m_3}   E_{\alpha+\beta, \ell_3,m_3}\\
\hskip 1.truecm \text{if} \ \ \alpha+ \beta\in \Sigma\\
c_{\ell_1,m_1,\ell_2,m_2}{}^{\ell_3,m_3}  \alpha\cdot H_{\ell_3,m_3} +  (-1)^{m_1} m_1 k\delta_{\ell_1,\ell_2} \delta_{m_1 + m_2}   \\
\hskip 1.truecm \text{if} \ \ \alpha + \beta=0\\
0  \hskip .8truecm
\text{otherwise}
\end{array}
\right.\nn\\
\big[d, H^i_{\ell,m}\big]&=& m H^i_{\ell,m} \ , \ \
\big[d, E_{\alpha,\ell,m}\big]= m E_{\alpha,\ell,m} \nn \ ,
\eeqa
 The coefficients $\epsilon(\alpha,\beta)=-\epsilon(\beta,\alpha)$ characterise the Lie algebra $\mathfrak{g}$ (and are equal to $\pm 1$ if $\g$ is simply laced).

Thus, for the two-torus the Cartan subalgebra is generated by $\{H^i_{00}, i=1,\cdots,r, d_1,d_2,k_1,k_2\}$ and
$\widehat{\mathfrak g}(\mathbb T_2)$ is of rank $r +4$, whilst for the two-sphere, the Cartan subalgebra is generated by
 $\{H^i_{00}, i=1,\cdots,r, d,k\}$ and
$\widehat{\mathfrak g}(\mathbb S^2)$ is of rank $r +2$. With respect to these Cartan subalgebras, the root spaces reduce to
\beqa
\label{eq:R-T}
\begin{split}
  \mathfrak{g}_{(\alpha,m,n)} &= \big\{E_{\alpha,m,n} \big\}\ , \ \ \alpha \in\Sigma, m,n \in \mathbb Z\\
  \mathfrak{g}_{(0,m,n)} &= \big\{H^i_{m,n}\ , i=1,\cdots,r \big\}\ ,  m,n \in \mathbb Z
  \end{split}
\eeqa
for the two-torus, and to
\beqa
\label{eq:R-S}
\begin{split}
  \mathfrak{g}_{(\alpha,m)} &= \big\{E_{\alpha,\ell ,m} \ , \ell \ge |m| \big\}\ , \ \ \alpha \in\Sigma, m \in \mathbb Z\\
  \mathfrak{g}_{(0,m)} &= \big\{H^i_{\ell,m}\ , i=1,\cdots,r,  \ell \ge |m|  \big\}\ ,  m,n \in \mathbb Z
  \end{split}
\eeqa
for the two-sphere. We have respectively the obvious commutation relations for the two-torus
\beqa
\label{eq:RR-T}
\Big[\mathfrak{g}_{(\alpha,m,n)},\mathfrak{g}_{(\beta,p,q)} \Big]\subset \mathfrak{g}_{(\alpha+\beta,m+p,n+q)} \ ,
  \ \ \Big[\mathfrak{g}_{(\alpha,m,n)},\mathfrak{g}_{(0,p,q)} \Big]\subset \mathfrak{g}_{(\alpha,m+p,n+q)}\
\eeqa
and for the two-sphere
\beqa
\label{eq:RR-S}
\Big[\mathfrak{g}_{(\alpha,m)},\mathfrak{g}_{(\beta,p)} \Big]\subset \mathfrak{g}_{(\alpha+\beta,m+p)}\ , \ \
  \Big[\mathfrak{g}_{(\alpha,m)},\mathfrak{g}_{(0,p)} \Big]\subset \mathfrak{g}_{(\alpha,m+p)}\ .
  \eeqa

For both  the two-sphere and the two-torus we can define an ordering relation:
\beqa
(\alpha,m,n)> 0 &\Leftrightarrow& n>0\ \  \text{or}\ \  (n=0, m>0)\ \  \text{or} \ \ (m=n=0, \alpha>0)\nn\\
(0,m,n)> 0 &\Leftrightarrow& n>0 \ \  \text{or} \ \ (n=0, m>0) \ , \nn
\eeqa
and
\beqa
(\alpha,m)> 0 &\Leftrightarrow& m>0\ \   \text{or} \ \ (m=0, \alpha>0)\nn\\
(0,m)> 0 &\Leftrightarrow& m>0  , \nn
\eeqa
respectively,  a fact that allows us to  define a set of positive and negative roots.

Even if \eqref{eq:RR-T} and \eqref{eq:RR-S} look similar, these two relations are substantially different.
Indeed, in the case of the two-torus, the root-space $\mathfrak{g}_{(\alpha,n,m)}$ (resp.  $\mathfrak{g}_{(0,n,m)}$)
is one dimensional (resp. $r-$dimensional), but we cannot extract a system of simple roots \cite{RMM}. Whereas in
the case of the two-sphere, the root spaces \eqref{eq:R-S} are infinite dimensional (recall that $\ell \ge |m|$).
This means that we can introduce an infinite number of simple roots corresponding to all possible values of
$\ell \ge |m|$ in the root spaces in \eqref{eq:R-S}. This alternative is not satisfactory for a rank $r+2$ Lie algebra. Consequently, in the case of
the two-sphere, the commutation relations \eqref{eq:RR-S} depend also on the representation theory of $SO(3)$ through
the appearance of Clebsch-Gordon coefficients in the Lie brackets. The reason why the root-spaces are one-dimensional for the two-torus
and infinite dimensional for the two-sphere is simply related to the fact that for the Kac-Moody algebras associated
to the former, we have two (equal to the dimension of $\mathbb T_2$) Hermitian operators, while for the two-sphere, we have
only one Hermitian operator. These observations extend to  Kac-Moody algebras associated to the compact manifold
${\cal M}$ equal to $G_c$ or $G_c/H$.

This short analysis clearly shows that Kac-Moody algebras associated to compact manifolds ${\cal M} = G_c, G_c/H$ don't belong to the class of Kac-Moody algebras defined through Cartan matrices. Indeed, for the former, all roots are known, whilst this is not the case for the latter. Secondly, for the former we do not have a system of simple roots whereas the latter admit simple roots.

\subsection{Chevaley-Serre presentation of the Kac-Moody algebra associated to the two-torus}

Even if we can't find a system of simple roots for $\widehat{\mathfrak{g}}(\mathbb T_2)$, it is possible to have a Chevalley-Serre presentation (observe that a Chevalley-Serre presentation is not possible ${\mathbb S}^2$). There are in fact two equivalent ways, either using the simple roots of $\g$ or,  as a consequence of \eqref{eq:aff}, using the simple roots of $\widehat{\g}$ the affine extension of $\g$.

In the former case, the Cartan matrix of $\g$ is given by
 $A_{ij}= 2 (\alpha_{(i)} \cdot \alpha_{(j)})/(\alpha_{(i)} \cdot \alpha_{(i)})$. Define
\beqa
h_{imp}= \alpha_{(i)}\cdot  H_{mp}  \ , \ \
e^\pm{}^i_{mp}=  E_{\pm \alpha_{(i)} mp} \ , \ \ i=1,\cdots, r  \ . \nn
\eeqa
The
Chevalley-Serre relations,  with ad$(x)\cdot y=[y,x]$, are given by ($i,j=1,\cdots,r$ and $n,m,p,q \in \mathbb Z$)
\beqa
\label{eq:CSS}
\begin{split}
\big[k_1,h_{imn}\big]&=\big[k_1,e^\pm{}^i_{mn}\big]=
\big[k_2,h_{imn}\big]=\big[k_2,e^\pm{}^i_{mn}\big]=0\\
\big[h_{imp}, h_{jnq}\big]&= (k_1 m + k_2 p)\;\alpha_{(i)} \cdot \alpha_{(j)} \; \delta_{m+n} \delta_{p+q}\\
\big[h_{imn},e^\pm{}^j_{pq}\big]&= \pm \alpha_{(i)}\cdot \alpha_{(j)}\; e^\pm{}^j_{m+pn+q}\\
\big[e^+{}^i_{mn},e^-{}^j_{pq}\big]&= \delta^{ij} \Big(h_{im+pn+q}+ (k_1 m + k_2 n) \delta_{m+n} \delta_{p+q}\Big)\\
{\rm  ad}^{1-A_{ij}}(e^\pm{}^i_{mn})\cdot e^\pm{}^j_{pq}&=0
\end{split}
\eeqa

The simple roots of $\widehat{\g}$ are given by
$\hat\alpha_{(i)}, i=0,\cdots,r$   with $\hat \alpha_{(i)}=(\alpha_{(i)},0,0)$, $i=1,\cdots,r$, $\hat \alpha_{(0)}=(-\psi,0,1)$,
where  $\psi$ is the highest root of $\g$.   The corresponding Cartan matrix is denoted by $\hat A_{ij}, i,j=0,\cdots,r$,
with scalar product (see for example \cite{go})
\beqa
\hat \alpha_1 \cdot \hat \alpha_2=(\alpha_1,k_1,m_1)\cdot (\alpha_2,k_2,m_2)= \alpha_1 \cdot \alpha_2 + k_1 m_2 +k_2 m_1\ .  \nn
\eeqa
 Introduce also the generators associated to the simple roots of $\widehat {\g}$
\beqa
\begin{array}{lll}
\hat e^\pm{}^i_{m} = e^\pm{}^i_{0m}, & \hat h_{im} = h_{i0m}& i=1,\cdots,r\\
\hat e^\pm{}^0_{m} = E^{\mp \psi}_{\pm 1  m}, & \hat h_{0m} =(-\psi\cdot H +k_1m).&
\end{array}
\nn
\eeqa
 This determines a presentation of the algebra (with $i,j=0,\cdots,r$ and $m,p\in \mathbb Z$) as
\beqa
\label{eq:Caff}
\begin{split}
\big[k_2,\hat h_{im}\big]&=\big[k_2,\hat e^\pm{}^i_{m}\big]={\rm ad}^{1-\hat A_{ij}}(\hat e^\pm{}^i_{m})\cdot \hat e^\pm{}^j_{p}=0\\
\big[\hat h_{im}, \hat h_{jn}\big]&= k_2 m \delta_{m+n} \;\hat\beta_{(i)} \cdot \hat \beta_{(j)},\quad
\big[\hat h_{im},\hat e^\pm{}^j_{p}\big]= \pm \hat \alpha_{(i)}\cdot \alpha_{(i)}\; \hat e^\pm{}^j_{ m+p} \\
\big[\hat e^+{}^i_{m},\hat e^-{}^j_{n}\big]&= \delta^{ij}\big( \hat h_{im+p} + k_2 m \delta_{m+n}\big)\\
\end{split}
\eeqa

The presentations \eqref{eq:CSS} and \eqref{eq:Caff} lead to the same algebra. Indeed, using the last relation in \eqref{eq:CSS}, we construct the operators
$e^\mp_{\psi, \pm 1, m} \equiv \hat e^\pm_{0m}$ associated to the highest root
$\psi$ and then define $\hat h_{0m}=[\hat e^+_{0m},\hat e^-_{00}]$. Thus, \eqref{eq:CSS} leads to \eqref{eq:Caff} with $\hat e^\pm_{0m}, \hat h_{0m}$ and $\hat e^\pm_{im} = e^\pm_{i0m}, \hat h_{im}= h_{i0m}$.
Conversely, we set $e^\pm_{i0m}=\hat e^\pm_{im},h_{i0m}= \hat h_{im}  $ and the action of the operators associated to the root $\hat \beta_{(0)}$ leads to the operators $e^\pm_{ipm}, h_{ipm}$, as it is usual in affine Lie algebras. Thus  \eqref{eq:Caff} leads to \eqref{eq:CSS}. Even if the two presentations define the same algebra, they are very different, as the former uses the roots of $\g$ and the Cartan matrix $A$,  whilst the latter is based on the
roots of $\widehat \g$ and the Cartan matrix $\hat A$.  The fundamental difference resides in the fact that
the Cartan matrices $A$ and $\hat A$ have very different properties.  While $A$ is an invertible matrix, $\hat A$ has corank equal to one ({\it i.e.}, its kernel is one-dimensional).
Finally observe that the presentation \eqref{eq:Caff} involves the quantum numbers of $\g$ and the presentation
\eqref{eq:CSS} the quantum numbers of $\widehat{\g}$.

The two presentations presented in this section have been used on a different footing. In \cite{MRT}, the presentation
\eqref{eq:Caff} was used to construct a vertex realisation of the algebra \eqref{eq:torus}. In \cite{ferm, bos}, the presentation
\eqref{eq:CSS} was used to construct an explicit fermionic and bosonic realisation of the Kac-Moody algebra
of the two-torus (in these articles,  a fermionic and bosonic realisation of the Kac-Moody algebra of the two-sphere was also
 obtained).

\section{Kac-Moody and Virasoro algebras associated to the two-sphere and the two-torus}
We now show that it is  possible to reproduce directly the Lie algebras of Sec. \ref{sec:KM}, but also a Virasoro algebra associated to
the two-torus and the two-sphere, considering the usual Kac-Moody and Virasoro algebras.

The Virasoro and Kac-Moody algebras have a semi-direct sum structure
\beqa
\label{eq:modes}
{\rm Vir}\ltimes \hat{\g} = \{L_m,c, m\in \mathbb Z\} \ltimes
\{T_m^a, k, a=1, \cdots, \dim \g, m \in \mathbb Z\}
\eeqa
with Lie brackets:
\beqa
\begin{split}
\big[L_m,L_n\big]&=(m-n) L_{m+n} + \frac c{12}m(m^2-1) \delta_{m+n} \\
\big[T_{m_1}^{a_1}, T_{m_2}^{a_2} \big] &={\rm i} f^{a_1 a_2}{}_{a_3} T^{a_3}_{m_1+m_2} + k m_1 \delta^{a_1 a_2} \delta_{m_1+m_2} \  \\
\big[L_m,T_{n}^{a}\big] &=-n T^a_{m+n} \ .
\end{split}
\eeqa
Observe that the grading operator  $d$   is identified with $-L_{0}$.
The generators in \eqref{eq:modes} can be understood as mode expansions of a conserved quantity on the circle $\mathbb S^1$ (with $0\le \varphi<2\pi$ being a parameterisation of the circle):
\beqa
\label{eq:exp}
L(\varphi)=  \sum \limits_{m=-\infty}^{+ \infty} L_{m}  e^{-i n \varphi}\ ,   \ \
T^a(\varphi) =\sum \limits_{m=-\infty}^{+ \infty} T_{m}^a e^{-i m \varphi}
\eeqa

The basic idea using \eqref{eq:modes} to reproduce \eqref{eq:torus} and \eqref{eq:sphere} is to consider the embedding
$\mathbb S^1\subset \mathbb T_2$ and $\mathbb S^1\subset \mathbb S^2$ in a certain way.\\

Consider in the first place the case of the two-torus, which is structurally simpler. The two-torus is now parameterised by $0\le\theta, \varphi<2\pi$.
Assume firstly to have a  mode expansion  analogous to \eqref{eq:exp}, but on the two-torus, {\it i.e.}, the conserved quantities
write now as $L(\theta,\varphi)$ and $T^a(\theta,\varphi)$. Performing the $\varphi-$integration, the
 Virasoro and Kac-Moody generators
$T^a_m, L_m$ depend now  on the variable $\varphi$:
\beqa
\frac 1{2\pi}\int \text{d} \varphi  e^{im \theta} T^a(\theta,\varphi)=T_m^a(\theta) &=& \sum \limits_{n=-\infty}^{+ \infty} T_{m,n}^a e^{-i n \theta} \nn\\
\frac 1{2\pi}\int \text{d} \varphi  e^{im \theta} L(\theta,\varphi)=L_m(\theta)&=& \sum \limits_{n=-\infty}^{+ \infty} L_{m,n}  e^{-i n \theta}.  \nn\
\eeqa

If we assume the Lie brackets
\beqa
\big[L_m(\theta),L_n(\theta')\big]&=&(m-n) L_{m+n} (\theta)\delta( \theta - \theta')+
\frac c{12}m(m^2-1) \delta_{m+n}\delta( \theta-\theta') \nn\\
\big[T_{m_1}^{a_1}(\theta), T_{m_2}^{a_2}(\theta') \big] &=& {\rm i} f^{a_1 a_2}{}_{a_3} T^{a_3}_{m_1+m_2}(\theta)
\delta(\theta-\theta')+ k m_1 \delta^{a_1 a_2} \delta_{m_1+m_2}\delta(\theta-\theta') \  \nn\\
\big[L_m(\theta),T_{n}^{a}(\theta')\big] &=&-n  T^a_{m+n} (\theta)\delta(\theta-\theta')\ . \nn
\eeqa
Integration by  $\int \text{d} \theta e^{i n_1 \theta} \int \text{d} \theta' e^{i n_2\theta'}$ on both sides leads to
\beqa
\label{eq:KV-T2}
\big[L_{m_1,n_1},L_{m_2,n_2}\big] &=& (m_1-m_2) L_{m_1+m_2,n_1+n_2} + \frac {c}{12}m_1(m_1^2-1)\delta_{m_1+m_2}\delta_{n_1+n_2}\nn\\
\big[T^{a_1}_{m_1,n_1},T^{a_2}_{m_2,n_2}\big] &=& {\rm i} f^{a_1 a_2}{}_{a_3} T^{a_2}_{m_1+m_2,n_1+n_2} + k \delta^{ab} m_1 \delta_{m_1+m_2}\delta_{n_1+n_2}\\
\big[L_{m_1,n_1},T^{a}_{m_2,n_2}\big]&=&-m_2 T^a_{m_1+m_2,n_1+n_2} \ .\nn
\eeqa
The second line reproduced the algebra \eqref{eq:torus}, but with one central charge and one differential operator
$d_1=-L_{00}$. This algebra is denoted
${\rm Vir}(\mathbb S^1 \times \mathbb S^1) \ltimes \widehat g_{(1)}(\mathbb S^1 \times \mathbb S^1)$. The symbol
$\g_{(1)}$ indicates that we have only one central extension. This will be clarified in the sequel.\\

The analysis can be extended to the two-sphere as follows. Let $0\le \varphi<2\pi, 0\le \theta \le \pi$ be a parameterisation of the two-sphere.
Introduce the spherical harmonics
\beqa
\label{eq:Y}
Y_{\ell m}(\theta,\varphi)= \sqrt{\frac{(\ell -m)!}{(\ell+m)!}}P_{\ell m}(\cos \theta)e^{i m \varphi} = Q_{\ell,m}(\cos \theta) e^{i m \varphi},
\eeqa
where $P_{\ell m}$ are the associated Legendre  functions,
satisfying the orthonormality  property  for $ \ell_1,\ell_2 \ge |m| $:
\beqa
\label{eq:orth}
(Q_{\ell_1 m}, Q_{\ell_2,m}) = \frac 12 \int \limits_0^\pi \sin \theta \text{d} \theta Q_{\ell_2,m}(\cos\theta) Q_{\ell_1,m}(\cos\theta)=
\delta^{\ell_1}_{\ell_2} \ ,   \
\eeqa
As  done for the case of the two torus, we assume now that the conserved quantities on $\mathbb S^2$
are
\beqa
L(\theta,\varphi)&=&   \sum\limits_{m=-\infty}^{+ \infty} \Big( \sum \limits_{\ell = |m|}^{+ \infty} L_{\ell,m}  Q_{\ell,m}(\cos\theta)\Big)
                         e^{i m\varphi}
                         = \sum\limits_{m=-\infty}^{+ \infty} L_m(\theta) e^{i m \varphi} \ , \nn\\
 T^a(\theta,\varphi)&=&   \sum\limits_{m=-\infty}^{+ \infty} \Big( \sum \limits_{\ell = |m|}^{+ \infty} T^a_{\ell,m}  Q_{\ell,m}(\cos\theta)\Big)
                         e^{i m\varphi}
                         = \sum\limits_{m=-\infty}^{+ \infty} T^a_m(\theta) e^{i m \varphi} \ , \nn
                         \eeqa

We further assume that  $L_m(\theta), T^a_m(\theta)$ satisfy the relations

\beqa
\big[L_m(\theta),L_n(\theta')\big]&=&(m-n) L_{m+n} (\theta)\delta(\cos \theta -\cos \theta')\nn\\
&&+ \frac c{12}m(m^2-1) \delta_{m+n}\delta(\cos \theta -\cos \theta') \nn\\
\big[T_{m_1}^{a_1}(\theta), T_{m_2}^{a_2}(\theta') \big] &=& {\rm i} f^{a_1 a_2}{}_{a_3} T^{a_3}_{m_1+m_2}(\theta) \delta(\cos \theta -\cos \theta')\nn\\
&&+ k m_1 \delta^{a_1 a_2} \delta_{m_1+m_2}\delta(\cos \theta -\cos \theta') \  \nn\\
\big[L_m(\theta),T_{n}^{a}(\theta')\big] &=&-n  T^a_{m+n} (\theta)\delta(\cos \theta -\cos \theta')\ . \nn
\eeqa
Now,  using again the orthonormality relation \eqref{eq:orth}, we get
\beqa
\label{eq:KV-S2}
\big[L_{\ell_1,m_1},L_{\ell_2,m_2}\big]&=&(m-n) c_{\ell_1,m_1,\ell_2,m_2}{}^{\ell_3,m_1+m_2}  L_{\ell_3,m_1+m_2} \nn\\
&&+(-1)^{m_1} \frac c{12}m_1(m_1^2-1)
\delta_{m_1+m_2}\delta_{\ell_1\ell_2} \\
\big[T_{\ell_1,m_1}^{a_1}, T_{\ell_2,m_2}^{a_2}\big]&=&{\rm i} f^{a_1a_2}{}_{a_3}  c_{\ell_1,m_1,\ell_2,m_2}{}^{\ell_3,m_1+m_2} T^{a_3}_{\ell_3,m_1+m_2}\nn\\
&&
+(-1)^{m_1} k m_1 \delta_{m_1+m_2}\delta_{\ell_1\ell_2} \nn\\
\big[L_{\ell_1,m_1},T_{\ell_2,m_2}^{a_2}\big]&=&-m_2 c_{\ell_1,m_1,\ell_2,m_2}{}^{\ell_3,m_1+m_2} T^{a_3}_{\ell_3,m_1+m_2}\ , \nn
\eeqa
The second line reproduces the algebra in \eqref{eq:sphere} with differential operator $d=-L_{00}$.  This algebra is denoted by
${\rm Vir}(\mathbb S^2) \ltimes \widehat{\g}(\mathbb S^2)$. \\

This construction provides a natural way to obtain extensions of the Virasoro and Kac-Moody algebras on the two-torus and the two-sphere, respectively. The first line in \eqref{eq:KV-T2} and \eqref{eq:KV-S2} can be seen as a subalgebra of vector fields on
 either the two-torus  or the two-sphere. These algebras are centrally extended.
To extend centrally a Lie algebra is far from being a trivial task. In the case of the semi-direct structure above, the situation is even more difficult.
Note that our procedure leads naturally to centrally extended algebras.
Only for the two-torus, we have ${\rm Vir}(\mathbb S^2) \ltimes \widehat \g(\mathbb S^2)\subset {\rm Vir} \ltimes \widehat \g$.
For the study of centrally extended  algebras of vector fields on $\mathbb T_2$ and $\mathbb S^2$, see for instance \cite{FI}. The compatibility  condition of centrally extended algebras of vector field of the two-torus and the two-sphere and the corresponding Kac-Moody algebra has been studied in \cite{FRS,RS,RS2}.

The analysis of the representation theory of the algebras \eqref{eq:KM-S2} and \eqref{eq:KV-T2} deserves a deeper analysis. However,
our process could be a way to construct
unitary representations of
${\rm Vir}(\mathbb T_2) \ltimes \widehat g_{(1)}(\mathbb T_2)$ or
${\rm Vir}(\mathbb S^2) \ltimes \widehat g(\mathbb S^2)$
by means of representations of
$\text{Vir} \ltimes \g$.
This observation is in accordance with the property that we obtain only
one central charge and one Hermitian operator in \eqref{eq:KV-T2}, as
it was proved in \cite{RMM} that unitarity implies that only  one central charge is non-vanishing.

Finally,  we observe that the construction presented in this section has been used to obtain fermions and bosons realisations of the algebras
${\rm Vir}(\mathbb T_2) \ltimes \widehat g_{(1)}(\mathbb T_2)$ or
${\rm Vir}(\mathbb S^2) \ltimes \widehat g(\mathbb S^2)$ \cite{ferm,bos}.
\begin{acknowledgement}
RCS  acknowledges financial support by the research
grant PID2019-106802GB-I00/AEI/10.13039/501100011033 (AEI/ FEDER, UE).

\end{acknowledgement}
\end{document}